\documentclass[11pt]{article}

\usepackage{verbatim}
\usepackage{amsmath}
\usepackage{amsfonts}
\usepackage{amssymb}
\usepackage{bbm}
\usepackage{bbold}
\usepackage{latexsym}
\usepackage{lscape} 
\usepackage{hhline} 
\usepackage{epsfig}
\usepackage[usenames,dvipsnames]{color}
\usepackage{graphicx}
\usepackage{enumitem}
\usepackage[normalem]{ulem} 
\usepackage[table]{xcolor}
\usepackage{subfig}
\usepackage{multirow}
\usepackage{mathrsfs}
\usepackage{blkarray}
\usepackage{amscd}
\usepackage{cite}

\DeclareMathSymbol{\mg}{\mathrel}{symbols}{"1D}

\newcommand{\bes}{\begin{split}}
\newcommand{\ees}{\end{split}}
\renewcommand{\arraystretch}{1.5}

\newcommand{\beq}{\begin{equation}}
\newcommand{\eeq}{\end{equation}}
\newcommand{\barr}{\begin{array}}
\newcommand{\earr}{\end{array}}

\newcounter{oldcounter}

\newcommand{\Intr}{\mathbbm{Z}}

\newcommand{\ba}[2]{\[\begin{array}{#2}\label{#1}}
\newcommand{\ea}{\end{array}\]}
\newcommand{\be}{\begin{equation}}
\newcommand{\ee}{\end{equation}}
\newcommand{\bea}{\begin{eqnarray}}
\newcommand{\eea}{\end{eqnarray}}

\begin{document}

\thispagestyle{empty}

\begin{flushright}
LTH-1257\\ 
\end{flushright}
\vskip 1 cm
\begin{center}
{\Large {\bf 
Uncovering a Spinor--Vector Duality on a Resolved Orbifold
} 
}
\\[0pt]

\bigskip
\bigskip {\large
{\bf A.E.~Faraggi$^{a,}$}\footnote{
E-mail: alon.faraggi@liverpool.ac.uk},
{\bf S.~Groot Nibbelink$^{b,c,}$}\footnote{
E-mail: s.groot.nibbelink@hr.nl},
{\bf   M.~Hurtado Heredia$^{a,}$}\footnote{
E-mail: martin.hurtado@liv.ac.uk}
\bigskip }\\[0pt]
\vspace{0.23cm}
${}^a$ {\it 
Department of Mathematical Sciences, University of Liverpool, Liverpool L69 7ZL, UK 
 \\[1ex] } 
${}^b$ {\it 
Institute of Engineering and Applied Sciences, Rotterdam University of Applied Sciences, \\ 
G.J.\ de Jonghweg 4 - 6, 3015 GG Rotterdam, the Netherlands
 \\[1ex]}
${}^c$ {\it 
Research Centre Innovations in Care, Rotterdam University of Applied Sciences, \\ 
Postbus 25035, 3001 HA Rotterdam, the Netherlands
 } 
\\[1ex] 
\bigskip
\end{center}

\subsection*{\centering Abstract}

Spinor--vector dualities have been established in various exact string realisations like orbifold and free fermionic constructions. This paper aims to investigate possibility of having spinor--vector dualities on smooth geometries in the context of the heterotic string. As a concrete working example the resolution of the $T^4/\Intr_2$ orbifold is considered with an additional circle supporting a Wilson line, for which it is known that the underlying orbifold theory exhibits such a duality by switching on/off a generalised discrete torsion phase between the orbifold twist and the Wilson line. Depending on this phase complementary parts of the twisted sector orbifold states are projected out, so that  different blowup modes are available to generate the resolutions. As a consequence, not only the spectra of the dual pairs are different, but also the gauge groups are not identical making this duality less apparent on the blowup and thus presumably on smooth geometries in general.

\newpage 
\setcounter{page}{1}

\section{Introduction}
\label{sc:Introduction}

String theory provides a perturbatively consistent framework for the unification of gravity with the gauge interactions. The number of string theories in ten dimensions is relatively scarce, and includes five theories that are supersymmetric and eight that are not. These ten dimensional theories are connected to lower dimension theories by compactification of the extra dimensions on some internal space such that they are hidden from contemporary experimental observations, resulting in a plethora of vacua in lower dimensions. 

String vacua in lower dimensions are in general studied by using exact worldsheet constructions, as well as effective field theory target space tools that explore the low energy spectrum of string compactifications. Among the exact worldsheet tools we may list the toroidal orbifolds~\cite{Dixon:1985jw,Dixon:1986jc}, the free fermionic formulation~ \cite{Antoniadis1987a,Kawai1987,AB} and the interacting Conformal Field Theory constructions~\cite{Gepner:1987vz}. The effective field theory models typically are obtained as compactifications of ten or eleven dimensional supergravity on a complex Ricci flat internal manifold~\cite{Candelas:1985en}. In light of all these constructions it is expected, that there may exist symmetries that underly the entire space of vacua in lower dimensions. In particular, mirror symmetry may be considered as an example of such a symmetry between vacua in lower dimensions~\cite{Greene:1990ud,Candelas:1990rm}. Mirror symmetry is believed to be related to T--duality~\cite{Strominger:1996it}. In toroidal orbifold compactifications T--duality arises due to the exchange of the moduli of the internal six dimensional compactified manifold, see {\it e.g.}~\cite{giveon_94}. 

The fermionic realization of $\Intr_2\times \Intr_2$ orbifolds led to the observation of a new symmetry in the space of heterotic--string compactifications, dubbed {\em spinor--vector duality}: two models are mapped to each other  under the exchange of the total number of spinorial plus anti--spinorial representations and the total number
of vectorial representations of an underlying $SO(2N)$ GUT symmetry group. The spinor--vector duality was initially observed by simple counting~\cite{Faraggi2007b}, using the classification tools developed in~\cite{Gregori:1999ny} for type II string, and in~\cite{Faraggi:2004rq,Faraggi2007a,Faraggi2007b} for heterotic--strings with unbroken $SO(10)$ GUT symmetry. This duality was proven to arise due to exchange of discrete generalised GSO phases in the free fermionic formulation~\cite{Faraggi2007b,CatelinJullien:2008pc}. In a bosonic representation of the spinor--vector duality~\cite{Angelantonj:2010zj,Faraggi:2011aw} the map between the dual vacua results from an exchange of a generalized discrete torsion on $\Intr_2$ toroidal orbifolds. Generalisation to compactifications with interacting intermal
CFTs was discussed in ref. \cite{Athanasopoulos:2014wha}.

In this paper, inspired by the spinor--vector duality, we would like to explore the existence of similar symmetries in compactifications of heterotic string theory on smooth Calabi--Yau manifolds with vector bundles. As a guideline for this exploration we start with  orbifold models discussed in~\cite{Faraggi:2011aw} on $T^4/\Intr_2\times S^1$ with a Wilson line on the additional circle. We then consider the resolution of this orbifold to a smooth K3$ \times S^1$ realisation and investigate how this effects the spinor--vector duality
(for earlier references on heterotic and open strings on K3 or
one of its orbifold realisations, see e.g.~\cite{Harvey:1986bf,Walton:1987bu}).
In particular, we show that this duality can still be realised, but in a more complicated guise. We take this effort as a first step to uncover spinor--vector dualities on smooth string compactifications in general.

\subsubsection*{Outline}

Section~\ref{sc:5DmodelWL} first recalls the description of the $T^4/\Intr_2$ orbifold of the heterotic $E_8\times E_8$ string. After that an additional circle is considered with a Wilson line. The effect of switching on a generalised torsion between the orbifold action and the Wilson line concludes this section. 

Section~\ref{sc:LineBundleResWL} describes some properties of the resolution of the $T^4/\Intr_2$ orbifold. In particular line bundle gauge backgrounds are introduced and the multiplicity operator is given to compute the full massless spectrum in six dimensions. 

The effect of the Wilson line is discussed next. The simplest case is the situation without the torsion phase switched on as the resulting five dimensional spectrum can just be analysed by field theory techniques. Since it is unclear how to switch on the generalised torsion phase between the orbifold twist and the Wilson line on the smooth side, an educated guess is made for its effects based on the results of the previous section on the orbifold theory with an additional circle. The effect of the Wilson line with torsion is that the twisted states which were used as blowup modes are kicked out and the resulting model seems to be inconsistent. This may be overcome by selecting other blowup modes which are kept by the Wilson line projection modified by the generalised torsion. Possible consequences of this for the spinor--vector duality conclude this section. 

The conclusion Section~\ref{sc:Conclusion} summarises the results obtained in the paper  and is completed by a short outlook on future directions.

%
%

\section{Five Dimensional $\boldsymbol{T^4/\Intr_2\times S^1}$ Model with Wilson Line }
\label{sc:5DmodelWL}

In this section a (very similar) orbifold model will be presented as studied in~\cite{Faraggi:2011aw}. There the orbifold $T^4/\Intr_2\times T^2$ with a Wilson line on one of the $S^1 \subset T^2$ was considered. The resulting models exhibit a spinor--vector duality induced by switching on/off a generalised GSO phase between the orbifold twist and the Wilson line: For one choice of the discrete torsion, the zero modes of the untwisted torus in the $\mathcal{N}=2$ twisted sector are attached to the spinorial characters of the GUT group, whereas for the other choice they are attached to the vectorial character. 

Since the second circle was just a spectator in the discussion of the spinor--vector duality in~\cite{Faraggi:2011aw}, it is omitted here for clarity, so that the focus is on the five dimensional geometry $T^4/\Intr_2\times S^1$ with a Wilson line on the circle. To demonstrate the effect the possible torsion phase between the orbifold twist and the Wilson line, first the theory on the orbifold $T^4/\Intr_2$ is recalled. For simplicity the orbifold standard embedding is chosen for the computation of the six dimensional massless states.  After that the Wilson line projections without or with torsion are taken into account to determine the resulting five dimensional spectra.

\subsection{Spectrum on $\boldsymbol{T^4/\Intr_2}$ in the Orbifold Standard Embedding}

This section begins with an introduction to the heterotic $E_8\times E_8$ string on the orbifold $T^4/\Intr_2$ using the orbifold standard embedding. The material here is standard and may be found {\em e.g.}\ in~\cite{Dixon:1985jw,Dixon:1986jc}; the notation used here follows~\cite{Ploger:2007iq}. The orbifold modular invariance condition
\begin{equation}
V^2 - v^2 \equiv 0~, 
\end{equation} 
is trivially solved by taking the non--zero entries of the twist and the gauge embedding identical: 
\begin{equation}
v = (\tfrac 12^2, 0^2)~, 
\qquad 
V = (\tfrac 12^2, 0^6)(0^8) 
\end{equation} 
The massless spectrum in six dimensions on the orbifold is determined by setting the left--moving mass 
\begin{equation}
0= M_R^2 = p_\text{sh}^2 -\tfrac 12 + \delta c~, 
\qquad 
p_\text{sh} = p + k\, v~, 
\end{equation}
and right--moving mass
\begin{equation}
0 = M_L^2 = P_\text{sh}^2 - 1 + \delta_c + \omega_i\, N_i + \overline{\omega}_i\, \overline{N}_i~, 
\qquad
P_\text{sh} = P + k\, V~, 
\end{equation} 
 to zero. Here $k = 0$ labels the untwisted sector and $k=1$ the twisted sector. The momenta $p$ and $P$ are taken from the lattices: 
\begin{equation}
p \in \Lambda_{SO(8)}~, 
\qquad 
P \in \Lambda_{E_8\times E_8}~. 
\end{equation}
Furthermore, the following notation is introduced: 
\begin{equation} 
\delta c = \tfrac 12\sum_i \omega_i (1- \omega_i)~, 
\qquad 
\omega_i \equiv k\, v_i~, 
\qquad 
\overline{\omega}_i \equiv - k\, v_i~, 
\end{equation} 
where $\equiv$ means equal up to integers, such that $0 < \omega_i, \overline{\omega}_i \leq 1$. Concretely, for the $\Intr_2$ orbifold at hand, this reduces to: 
$\omega_i = \overline{\omega}_i = 1$,  $\delta c = 0$ for $k=0$ and  
$\omega_i = \overline{\omega}_i = \tfrac 12$,  $\delta c = \tfrac 14$ for $k=1$.  

The gauge group in six dimensions is determined by
\begin{equation}
P^2 = 2~, 
\qquad 
V\cdot P \equiv 0~. 
\end{equation}
These come as bosons (gauge fields) with $p=\pm (0^2,\underline{1,0})$ and spinors (gauginos) with $p=(\underline{\tfrac 12, -\tfrac 12}, \underline{-\tfrac12, \tfrac 12})$. Here the underline indicates that all possible permutations are to be considered as well. The untwisted charged matter is characterized by: 
\begin{equation}
V \cdot P - v\cdot p \equiv 0~, 
\end{equation}
with $N_i = \overline{N}_i = 0$. 
In addition, there are uncharged untwisted matter with $P=0$ and one $N_i, \overline{N}_i$ equal to 1 and the rest zero. These states come as bosons with $p=\pm(\underline{1,0},0^2)$ and spinors of the opposite chirality (hyperinos) $p=\pm(\tfrac 12^2, \pm\tfrac 12^2)$. 
The twisted matter comes in multiples of 16 due to the fact that there are 16 fixed points. The right-moving momentum is fixed to:
\begin{equation}
p_\text{sh}^2 = \tfrac 12~. 
\end{equation}
Hence, we can only have the bosonic states $p_\text{sh} = (\underline{\tfrac 12,-\tfrac 12},0^2) = v + p$ with $p=(\underline{0,-1},0^2)$ and fermionic states $p_\text{sh} = (0,0,\pm\tfrac 12^2) =v +p$ with $p=(-\tfrac 12^2,\pm\tfrac 12^2)$. Their representation with respect to the gauge group are determined by 
\begin{equation}
P_\text{sh}^2 = \tfrac 32 - \tfrac 12 \sum_i (N_i + \overline{N}_i)~. 
\end{equation}
The solutions of this mass equation are well--known~\cite{Ibanez:1997td,Honecker:2006qz} and are summarized in Table~\ref{tb:T4Z2StEmbSpec} for later convenience. 

\begin{table} 
\[ \renewcommand{\arraystretch}{1.6}
\begin{array}{|c|c|c|c|c|} 
\hline 
P_\text{sh}  & W\cdot P_\text{sh}\equiv & SU(2)\times E_7 \times E_8' & SU(2)_1\times SU(2)_2 \times SO(12) \times SO(16)' 
\\ \hline\hline   
\pm(1^2,0^6)(0^8)  & 0 & SU(2)~\text{gauge}  & SU(2)_1~\text{gauge}
\\  \hline
\pm(0^2, \underline{\pm1^2,0^4})(0^8) & 0 & E_7~\text{gauge} & SO(12)~\text{gauge}
\\ 
(\underline{1,-1},0^6)(0^8) & 0 &  & SU(2)_2~\text{gauge}
\\ 
( \underline{\tfrac 12, -\tfrac 12}, \underline{-\tfrac 12^o, \tfrac 12^{6-o}})(0^8)  & \tfrac 12 &  & (1,2,32)(1)~\text{gauge}
\\ \hline
(0^8,0^8) & 0 & 4\, (1,1)(1) & 4\, (1,1,1)(1) 
\\ \hline 
\pm(\underline{1,0},\underline{\pm1,0^5})(0^8)  & 0 & (2,56)(1)  & (2,2,12)(1) 
\\  
\pm(\tfrac 12^2, \underline{-\tfrac 12^e, \tfrac 12^{6-e}})(0^8)  & \tfrac 12 &  & (2,1,32)(1)
\\ \hline  
(\underline{\tfrac 12,-\tfrac 12},\underline{\pm1,0^5})(0^8) & 0  & 16\,\tfrac 12(1,56)(1) & 16\,\tfrac 12(1,2,12)(1)
\\  
\pm ( 0^2, \underline{-\tfrac 12^e, \tfrac 12^{6-e}})(0^8) & \tfrac 12 &  & 16\,\tfrac 12(1,1,32)(1)
\\  \hline  
\pm(\tfrac 12^2,0^6)(0^8) & 0 & 32\,(2,1)(1)  & 32\, (2,1,1)(1) 
\\ \hline\hline  
\pm (0^8)(\underline{\pm1^2,0^6}) & 0 &  E_8~\text{gauge} & SO(16)~\text{gauge} 
\\
\pm (0^8) ( \underline{-\tfrac 12^e, \tfrac 12^{8-e}}) & \tfrac 12 &  & (1,1,1)(128)~\text{gauge}  
\\  \hline 
\end{array}
\]
\caption{\label{tb:T4Z2StEmbSpec}
This table gives the weights of the massless states on the $T^4/\Intr_2$ orbifold. In addition, the eigenvalue $W\cdot P_\text{sh}$ and the resulting branching of this spectrum due to the Wilson line $W$ is indicated. (The underline indicates that all permutations are to be considered and $o$ and $e$ go over all odd and even numbers, respectively, such that the powers never go negative.) The matter multiplets are hyper multiplets or half--hyper multiplets ({\em i.e.}\ hyper multiplets with a reality condition imposed), the latter are indicated by the $\tfrac 12$ in front of the states. 
}
\end{table}

\subsection{Orbifold with Wilson line on an Additional Circle -- No Torsion}
\label{sc:OrbiWilsonLinNoTorsion}

Next, the model is compactified further down to five dimensions by a discrete $\Intr_2$ Wilson line given by $W=(0^7,1)(0^7,1)$ on an additional circle $S^1$. In this subsection the option of adding a torsion phase between the orbifold action and the Wilson line is ignored; this will be considered in the next subsection. The projection conditions are determined by the requirement~\cite{Ploger:2007iq}:
\begin{equation} \label{eq:PhaseProjection} 
e^{2\pi i (V_{h'} \cdot P_\text{sh} - v_{h'}\cdot p_\text{sh} + v_{h'}\cdot(N - \overline{N}))}\cdot e^{2\pi i \tfrac 12 (V_{h'}\cdot V_h - v_{h'}\cdot v_h)} \stackrel{!}{=} 1~, 
\end{equation}
where $V_h = k\, V+ n\, W$, $v_h = k\, v$, $P_\text{sh} = P + V_h$ and $p_\text{sh} = p+v_h$. 
The first factor can be understood in field theory while the second factor is the vacuum phase of the string. 
For the choice 
\begin{equation} \label{eq:WilsonLine} 
v = (\tfrac 12^2, 0^2)~, 
\quad 
V = (\tfrac 12^2, 0^6)(0^8)~, 
\quad 
W=(0^7,1)(0^7,1)~, 
\end{equation}
the vacuum phase is trivial: 
%
 \(
 \tfrac 12 (V_{h'}\cdot V_h - v_{h'}\cdot v_h) 
 \equiv 0\,. 
 \nonumber 
\)
%
The first phase in \eqref{eq:PhaseProjection} leads to the orbifold projection: 
\begin{equation}
V\cdot P - v\cdot p + (N_i - \overline{N}_i)v_i \equiv 0
\end{equation}
and the projection due to the Wilson line: 
\begin{equation} \label{eq:WilsonProjectionNoTorsion} 
W\cdot P \equiv 0~. 
\end{equation}
Here we are using the fact that we are considering a general point in the moduli space,
hence no winding mode states accidentally become massless. Consequently,
the Wilson line reduces the gauge group to 
\begin{equation}
SU(2)_1 \times SU(2)_2 \times SO(12) \times SO(16)'
\end{equation}
with the spectrum in five dimensions: 
\begin{equation} \label{eq:5DSpectrumNoTorsion} 
(2,2,12)(1) + 16\, \tfrac 12(1,2,12)(1) + 32\, (2,1,1)(1) + 4 (1,1,1)(1)~. 
\end{equation}
Note, in particular, that this spectrum does not contain any spinorial representation of $SO(12)$.

\subsection{Orbifold with Wilson line on an Additional Circle  -- With Torsion}

In this subsection the same compactification on $S^1$ with the Wilson line \eqref{eq:WilsonLine} is investigated, but now the option of switching on a torsion phase between de orbifold action and the Wilson line is considered~\cite{Ploger:2007iq}: 
\begin{equation} \label{eq:Torsion} 
T = (-1)^{\epsilon(k n' - k' n)}~, 
\end{equation}
where $k=0, 1$ labels the untwisted ($k=0$) and the twisted ($k=1$) sectors and $n=0,1$ the $\Intr_2$ Wilson line sectors. The primed versions define the orbifold and Wilson line projections. The torsion phase is switched on and off for $\epsilon=1$ and $0$, respectively.\footnote{Note that is a generalisation of discrete torsion considered in~\cite{Vafa:1986wx,Vafa:1994}, which is between two orbifold twists.}
The projection conditions \eqref{eq:PhaseProjection} are then modified to: 
\begin{equation}
e^{2\pi i (V_{h'} \cdot P_\text{sh} - v_{h'}\cdot p_\text{sh} + v_{h'}\cdot(N - \overline{N}))}\cdot e^{2\pi i \tfrac 12 (V_{h'}\cdot V_h - v_{h'}\cdot v_h)} 
\cdot 
e^{2\pi i\, \tfrac 12(kn' -k'n)}
\stackrel{!}{=} 1~, 
\end{equation}
Away from special points in moduli space, the winding modes with $n=1$ will be massive and hence do not affect the massless spectrum and will therefore be ignored. Thus when the torsion phase is switched on it only modifies the Wilson line projection \eqref{eq:WilsonProjectionNoTorsion}  to: 
\begin{equation} \label{eq:WilsonProjectionWithTorsion} 
W\cdot P_{sh} \equiv \tfrac 12\,k~: 
\end{equation}
Thus, for the untwisted states the projection is the same as without torsion, for the twisted states things change: The twisted states that were projected out before are kept with the torsion phase and vise versa. Thus, in particular,  the gauge group remains the same 
\begin{equation}
SU(2)_1 \times SU(2)_2 \times SO(12) \times SO(16)'
\end{equation}
but the spectrum in five dimensions changes to: 
\begin{equation} \label{eq:5DSpectrumTorsion} 
(2,2,12)(1) + 16\, \tfrac 12(1,1,32)(1) + 4 (1,1,1)(1)~. 
\end{equation}
As compared to the spectrum \eqref{eq:5DSpectrumNoTorsion} the 16 vectorial half--hyper multiplets $(1,1,12)(1)$ have been replaced by 16 spinorial half--hyper multiplets $(1,1,32)(1)$ and the 32 doublets $(2,1,1)(1)$ have been removed all together. Switching the torsion \eqref{eq:Torsion} on or off thus induces a spinor--vector duality between the two Wilson line models considered in these two subsections.

\section{Line Bundle Resolutions of $\boldsymbol{T^4/\Intr_2\times S^1}$ with Wilson Lines}
\label{sc:LineBundleResWL}

\subsection{Geometry of the $\boldsymbol{T^4/\Intr_2}$ Resolution}

The techniques to determine resolutions of toroidal orbifolds have been well--studied~\cite{Nahm:1999ps,Wendland:2001ks,Denef:2004dm}; here in particular the methods exploited in~\cite{Lust:2006zh} are used. The resolution of the $T^4/\Intr_2$ orbifold can be described with four inherited divisors $R_1, R_1', R_2,R_2'$, eight ordinary divisors $D_{1,\alpha_3\alpha_4}$ and $D_{2,\alpha_1\alpha_2}$ and sixteen exceptional divisors $E_{\alpha}$\,, where $\alpha=(\alpha_1,\alpha_2,\alpha_3,\alpha_4)$ with $\alpha_i=0,1$ labels the sixteen isolated $\Intr_2$ singularities on the torus $T^4$. There are a number of linear relations among these divisors
\begin{equation}
R_1 \sim R_1' \sim 2\, D_{1,\alpha_3\alpha_4} + \sum_{\alpha_1,\alpha_2} E_{\alpha}~, 
\qquad 
R_2 \sim R_2' \sim 2\, D_{2,\alpha_1\alpha_2} + \sum_{\alpha_3,\alpha_4} E_{\alpha}~. 
\end{equation}
These relations show that ordinary divisors $D_{1,\alpha_3\alpha_4}$ and $D_{2,\alpha_1\alpha_2}$ and inherited divisors $R_1'$ and $R_2'$ may be replaced by inherited divisors $R_1, R_2$ and exceptional divisors $E_{\alpha}$. The non--vanishing intersection numbers of the remaining divisors may be summarised as: 
\begin{equation}
R_1 R_2 = 2~, 
\qquad 
E_{\alpha} E_{\beta} = -2\,\delta_{\alpha\beta}~. 
\end{equation} 
The total Chern class may be represented as 
\begin{equation}
c = (1-R_1)(1-R_1')(1-R_2)(1-R_2') 
\prod_{\alpha_3,\alpha_4} (1 +D_{1,\alpha_3\alpha_4}) 
\prod_{\alpha_1,\alpha_2} (1 +D_{2,\alpha_1\alpha_2}) 
\prod_{\alpha} (1 +E_{\alpha})~. 
\end{equation}
Expanding this to first and second order gives 
\begin{equation}
c_1 = 0~,
\qquad 
c_2 = 24~. 
\end{equation} 
The first signifies that this resolution is a four dimensional K3 surface with Euler number 24 as the second Chern class $c_2$ indicates.

\subsection{Line Bundles on the $\boldsymbol{T^4/\Intr_2}$ Resolution} 

For orbifold resolution models it is generically assumed that the gauge flux is located on the exceptional divisors only. Hence, the line bundle background encoded by an anti--Hermitian Abelian gauge field strength two--form $\mathcal{F}_2$ given by~\cite{Nibbelink:2007rd,Nibbelink:2007pn,Nibbelink:2009sp,Blaszczyk:2010db}: 
\begin{equation}
\dfrac{\mathcal{F}_2}{2\pi} = \sum_\alpha E_\alpha\, \mathsf{H}_\alpha~, 
\quad 
\mathsf{H}_\alpha =  V_\alpha^I\, \mathsf{H}_I~,
\end{equation}
where the sum over the Cartan generators labelled by $I$ is implied. The Cartan generators $\mathsf{H}_I$ of $E_8\times E_8$ are normalized such that $\text{tr}\, \mathsf{H}_I \mathsf{H}_J = \delta_{IJ}$\,. The embedding of the line bundle background is therefore characterized by sixteen component line bundle (embedding) vectors $V_\alpha = (V_\alpha^I)$. (For translations to other characterizations see e.g.~\cite{Nibbelink:2015ixa}.) Often it is convenient to split the line bundle vectors in contributions in the first and second $E_8$ as: $V_\alpha = (\vec{V}_\alpha)(\vec{V}_\alpha')$ where $\vec{V}_\alpha$ and $\vec{V}'_\alpha$ both have 8 entries. 

The fundamental consistency requirement of such backgrounds is determined from the integrated Bianchi identity $\text{tr}(\mathcal{F}_2^2) - \text{tr}(\mathcal{R}_2^2) = 0$\,. On this resolution it can be cast in the form: 
\begin{equation}
 \sum_\alpha V_\alpha^2 = 24~. 
\end{equation}

The six dimensional spectrum of the line bundle model can be computed using multiplicity operator~\cite{Nibbelink:2007rd}: 
\begin{equation}
\mathsf{N} = -\int 
\Big\{ \dfrac 12\, \Big(\dfrac{\mathcal{F}}{2\pi}\Big)^2 - \dfrac 1{24}\, \text{tr} \Big( \dfrac{\mathcal{R}}{2\pi} \Big)^2 \Big\}
= \sum_\alpha \mathsf{H}_\alpha^2 - 2~. 
\end{equation}
This operator counts the number of fermions in a given representation. The sign of this operator may be positive or negative and is determined by the six dimensional chirality of the underlying fermionic states: It equals $-2$ on gaugino states; the multiplicity operator directly identifies the gauge group unbroken by the line bundle background. The multiplicity operator $\mathsf{N}$ is positive on hyperinos as they have the opposite chirality as gauginos  in six dimensions. Hence, if positive, it counts the number of hyper multiplets in a given representation of the gauge group.

\subsection{Line Bundle Model with Vectorial Blowup Modes}

Consider the resolution model with line bundle vectors 
\begin{equation} \label{eq:LineBundle} 
V_\alpha  = (\tfrac 12, -\tfrac 12, 1, 0^5)(0^8)~, 
\end{equation} 
at all sixteen fixed points. Since the same line bundle is chosen on all sixteen exceptional divisors, the multiplicity operator simplifies to
\begin{equation} \label{eq:MultOpAllEqual} 
\mathsf{N} = 16\, \mathsf{H}_V^2 - 2~.
\end{equation} 
Using this operator the multiplicities of the $E_8\times E_8'$ roots can be computed. The resulting spectrum is given in Table~\ref{tb:ResT4Z2StEmbSpec}. The assignment of untwisted and twisted states in this table is done by comparing with the untwisted states on the orbifold which can be understood as from field theoretical orbifolding of the $E_8\times E_8'$ ten-dimensional gauge multiplet. Since $T^4/\Intr_2$ has 16 fixed points and all fixed points (and their blowups) are treated identically, multiples of 16 are required.

\begin{table}
\[ \renewcommand{\arraystretch}{1.6}
\begin{array}{|c|c|c|c|c|} 
\hline 
\text{weight} & \mathsf{H}_V^2  & \mathsf{N} &  SU(2)\times E_6 \times E_8' & SU(2) \times SO(10) \times SO(16)' 
\\ \hline\hline   
\pm(1,1,0,0^5)(0^8) & 0 & -2_U & SU(2)~\text{gauge}  & SU(2)~\text{gauge}
\\  \hline
\pm(0,0,0, \underline{\pm1^2,0^3})(0,^8) & 0 & -2_U & E_6~\text{gauge} & SO(10)~\text{gauge}
\\ 
\pm ( \tfrac 12, -\tfrac 12, -\tfrac 12, \underline{-\tfrac 12^e, \tfrac 12^{5-e}})(0^8)  &  &  &  & (1,16)(1)~\text{gauge}
\\ \hline
\pm(1,0,-1,0^5)(0^8) & \tfrac 14 & 2_U & (2,27)(1)  & (2,1)(1) 
\\  
\pm(0,-1,-1,0^5)(0^8) & & &   & 
\\  
\pm(\underline{1,0},0, \underline{\pm1,0^4})(0^8) &  &  &  & (2,10)(1)
\\  
\pm (\pm\tfrac 12^2, -\tfrac 12, \underline{-\tfrac 12^o, \tfrac 12^{5-o}})(0^8)  &  &  &  & (2,16)(1)
\\ \hline  
\pm(1,-1,0,0^5)(0^8) & 1 & 14 = 16_T -2_U & (1,27)(1) & (1,1)(1) 
\\  
\pm(0,0,1, \underline{\pm1,0^4})(0^8) &  &  &  & (1,10)(1) 
\\  
\pm ( \tfrac 12, -\tfrac 12, \tfrac 12, \underline{-\tfrac 12^o, \tfrac 12^{5-o}})(0^8) &  &  &  & (1,16)(1)
\\  \hline  
\pm(1,0,1,0^5)(0,0^7) & \tfrac 94 & 34=32_T+2_U & (2,1)(1)  & (2,1)(1) 
\\  
\pm(0,-1,1,0^5)(0,0^7) & & &   & 
\\ \hline\hline  
\pm (0^8)(\underline{\pm1^2,0^6}) & 0 & -2_U & E_8~\text{gauge} & SO(16)~\text{gauge} 
\\
\pm (0^8) ( \underline{-\tfrac 12^e, \tfrac 12^{8-e}}) &  &  &  & (1,1)(128)~\text{gauge}  
\\  \hline 
\end{array}
\]
\caption{\label{tb:ResT4Z2StEmbSpec}
The multiplicities of the $E_8\times E_8$ roots are indicated for the resolution model generated by identical vectorial blowup modes at all sixteen exceptional divisors. The states with a positive or a negative multiplicity form hyper or vector multiplets. The subscripts $U$ and $T$ indicate how these numbers can be used to interpret the corresponding states as untwisted or twisted, respectively. The final column gives the spectrum branched by the Wilson line. }
\end{table}

\subsubsection*{Matching with the Orbifold Spectrum}

The above resolution model can be understood as a blowup of the orbifold standard embedding model. The techniques to understand the relations between the orbifold and resolutions spectra were discussed in {\em e.g.}~\cite{GrootNibbelink:2007ew,Nibbelink:2008tv,Nibbelink:2009sp,Blaszczyk:2010db}. The choice of the line bundle vectors as \eqref{eq:LineBundle} can be interpreted as using the identical blowup modes with this shifted momenta 
\begin{equation} \label{eq:IdBLW} 
P_{\text{sh},\alpha} = V_\alpha  = V + P~, 
\quad 
V= (\tfrac 12,\tfrac 12, 0,0^5)(0^8)
\quad\text{and}\quad 
P= (0,-1,1,0^5)(0^8)
\end{equation}
at all sixteen fixed points. These shifted momenta identify the sixteen blowup modes to lie inside the $(1,2,12)(1) \subset (1,56)(1)$ half--hyper multiplets given in Table~\ref{tb:T4Z2StEmbSpec}. Switching on these blowup modes leads to the symmetry breaking: 
\begin{equation} \label{eq:WilsonLineGaugeBreaking} 
SU(2) \times E_7 \times E_8' \rightarrow SU(2) \times E_6 \times E_8' 
\end{equation}
In this process precisely the roots $\pm(1,1,0,0^5)(0^8)$\,, $\pm(0,0,1,\underline{\pm1,0^4})$ and 
$\pm(\tfrac 12,\tfrac 12,\tfrac 12, \underline{-\tfrac12^o,\tfrac12^{5-o}})$ of the $(1,27)(1)$ are broken. This corresponds to the computation $14 = 16_T - 2_U$, which can be understood as the super--Higgs effect where certain twisted states are "eaten" to form massive vector multiples. These are states that arise from the sixteen half--hyper multiplets $(1,56)(1)$. Under the symmetry breaking this branches to 
\begin{equation}
\tfrac 12(1,56)(1) \rightarrow (1,27)(1) + (1,1)(1)~, 
\end{equation}
where the states $(1,1)(1)$ can be identified as the blowup modes (BLW). On the resolution they are reinterpreted as sixteen model dependent axions, which do not contribute to the multiplicity operator~\cite{GrootNibbelink:2007ew}.

The remaining fourteen states $(1,27)(1)$ after the Higgsing undergo a field redefinition when moving from the hyper multiplets on the orbifold to the states on the resolution: 
\begin{equation}
(1,27)(1)_\text{RES} = \text{BLW}^{-1}\cdot (1,27)(1)_\text{ORB} 
\end{equation}
Here the subscripts $\text{RES}$ and $\text{ORB}$ indicate whether the states are part of the resolution or orbifold description, respectively. Indeed, the corresponding weights can be matched exactly via:
\begin{equation}\renewcommand{\arraystretch}{1.6}
\begin{array}{ccccc}  
 \pm(0,0,-1,\underline{\pm1,0^4})(0^8)
 &  =  & 
 \pm \big[ -(\tfrac 12,-\tfrac 12, 1,0^5)(0^8) & + &  (\tfrac 12,-\tfrac 12, 0, \underline{\pm1,0^4}) \big]   
 \\
 \pm (-\tfrac 12,-\tfrac 12,-\tfrac 12,\underline{-\tfrac 12^o,\tfrac 12^{5-o}})(0^8) 
 &  =  & 
 \pm \big[ -(\tfrac 12,-\tfrac 12, 1,0^5)(0^8) & + & (0,0,\tfrac 12, \underline{-\tfrac 12^o,\tfrac 12^{5-o}}) \big]   
 \\
 \pm(-1,1,0,0^5)(0^8)
  &  =  &
  \pm \big[ -(\tfrac 12,-\tfrac 12, 1,0^5)(0^8) & + &  (-\tfrac 12,\tfrac 12, 1, 0^5) \big] 
\end{array}
\end{equation}
A similar field redefinition relate the doublet states on the orbifold to those on the resolutions: 
\begin{equation}
(2,1)(1)_\text{RES} = \text{BLW}\cdot (2,1)(1)_\text{ORB}~, 
\end{equation}
or in terms of the corresponding weights: 
\begin{equation}
\begin{array}{rcl} 
 \pm (1,0,1,0^5)(0^8)   &=& \pm\big[(\tfrac 12,-\tfrac 12, 1,0^5)(0^8) + (\tfrac 12^2, 0, 0^5) \big]
 \\[1ex]  
  \pm (0,-1,1,0^5)(0^8)   &=& \pm\big[(\tfrac 12,-\tfrac 12, 1,0^5)(0^8) + (-\tfrac 12^2, 0, 0^5)\big]
 \end{array}
\end{equation}
Hence, using these field redefinitions the descriptions on the orbifold and on the resolutions agree on the level of the weights showing that the matching between the orbifold and resolved descriptions is complete.

\subsection{Wilson Line Projected Vectorial Blowup Model}
 
 Next the consequences of the Wilson line on the additional circle is investigated in the resolved geometry. There are two cases to be considered depending on whether a generalisation of the torsion phase~\eqref{eq:Torsion} has been switched on or not. On smooth geometries it is less clear how to implement the string torsion phases as the description starts from an effective field theory description in ten dimensions rather than the full one--loop partition function of string theory. For this reason the Wilson line projection conditions are strongly inspired by the conditions arising in the orbifold theory.

\subsubsection*{No Torsion}

The model is compactified further on a circle $S^1$ with a discrete Wilson line: 
\begin{equation}  
W = (0^7, 1)(0^7,1)~ 
\end{equation}
and the torsion phase \eqref{eq:Torsion} is switched off: $\epsilon=0$\,. The Wilson line projection condition on the resolution is assumed to take the form: 
\begin{equation} 
W\cdot P \equiv 0~,
\end{equation}
where $P$ are the weights listed in Table~\ref{tb:ResT4Z2StEmbSpec}. 
This directly follows from the orbifold Wilson line projection~\eqref{eq:WilsonProjectionNoTorsion}, since the difference between the $P_\text{sh}$ and $P$ is at most given by $V_\alpha$, but $W\cdot V_\alpha = 0$. The gauge group therefore becomes: 
\begin{equation}
SU(2)\times SO(10) \times SO(16)'
\end{equation}
and the 5D spectrum: 
\begin{equation}
2\, (2,10)(1) + 36\, (2,1)(1) + 14\, (1,10)(1) + 14\, (1,1)(1)~.  
\end{equation} 
Notice the absence of any spinors of $SO(10)$ in this resolution model after the Wilson line projection has been implemented.

This spectrum can also be understood as the blowup of the five dimensional $T^4/\Intr_2 \times S^1$ model with the same Wilson line $W$ discussed in Subsection~\ref{sc:OrbiWilsonLinNoTorsion}. The blowup using the blowup modes~\eqref{eq:IdBLW} leads to the gauge symmetry breaking: 
\begin{equation}
SU(2)_1\times SU(2)_2 \times SO(12)\times SO(16)' \rightarrow
SU(2)_1 \times SO(10)\times SO(16)'~. 
\end{equation}
The broken weights are $(\underline{1,-1},0,0^5)(0^8)$ associated to $SU(2)_2$ and $\pm(0,0,1,\underline{\pm1,0^4})$ associated to $2\, (1,10)(1)$. 
The spectrum is branched as follows: 
\begin{equation}
\begin{array}{rcl}
(2,2,12)(1) &\rightarrow& 4\, (2,1)(1) + 2\,(2,10)(1)~, 
\\[2ex] 
16\, \tfrac 12(1,2,12)(1) &\rightarrow &16\, (1,10)(1) +32\, (1,1)(1)~, 
\\[2ex] 
32\, (2,1,1)(1) &\rightarrow&  32\, (2,1)(1)~. 
\end{array}
\end{equation}
The number of $(2,10)(1)$ immediately agree, so do the number of $(2,1)(1)$: $4+32=36$. 
Of the sixteen $(1,10)(1)$'s two are eaten to form massive $(1,10)(1)$ vector multiplets leaving fourteen states. Finally, sixteen of the 32 charged singlets $(1,1)(1)$ should be identified as blowup modes and hence appear as axions in the resolved theory. Furthermore, two singlets are eaten to make the $SU(2)_2$ weights heavy, leaving $32 - 16 - 2 = 14$ charged singlets in the spectrum.

\subsubsection*{With Torsion}

The description of the Wilson line on the additional circle with no torsion is fully self--consistent as was discussed above. On the contrary, switching the torsion~\eqref{eq:Torsion}, i.e.\ $\epsilon=1$, leads to a number of issues: 

First of all, it is not clear how to precisely implement the Wilson line projection in this case on the resolution.  On the orbifold the projection condition~\eqref{eq:WilsonProjectionWithTorsion} with torsion distinguishes between untwisted and twisted states. While on generic  smooth compactifications such a distinction is completely meaningless, for smooth models obtained as orbifold resolutions it is possible to make an assignment of "untwisted" and "twisted" states as indicated in Table~\ref{tb:ResT4Z2StEmbSpec} and Table~\ref{tb:ResT4Z2StEmbSpecSpinorial} based on intuition from and matching with the underlying orbifold theory. Hence, it is natural to assume that the discrete torsion modifies the projection condition analogously to the orbifold case.

Secondly, the blowup modes~\eqref{eq:IdBLW} used to generate the blowup model are projected out by the Wilson line when the torsion is switched on as the projection condition is modified to~\eqref{eq:WilsonProjectionNoTorsion}, since the blowup modes are twisted states with $k=1$\,. (Resolution models with bundle vectors that would be  associated with massive or projected out twisted states have been known in the literature but are not well--understood.)

A closely related issue is that there are states missing for the super--Higgs mechanism to be able to operate. On the orbifold the Wilson line would project the gauge group to $SU(2)\times SO(12) \times SO(16)'$. The blowup procedure leads to further breaking $SO(12) \rightarrow SO(10)$ hence two $(10)$'s of $SO(10)$ should form massive multiplets with $(10)$--plets as hyper multiplets. (The $14 = 16_T -2_H$ computation discussed below~\eqref{eq:WilsonLineGaugeBreaking}.)  But these twisted $(10)$-plets are projected out by the Wilson line when the torsion is switched on.

\subsection{Line Bundle Model with Spinorial Blowup Modes}

The main issue with the resolution model discussed just above is that the blowup modes which are supposed to generate the blowup are projected out by the Wilson line when the torsion is switched on. On the level of the orbifold this projection kicks out twisted vectorial states, including the blowup modes~\eqref{eq:IdBLW}, while keeping spinorial ones. As it is a choice which twisted states are used as blowup modes, it instructive to investigate spinorial blowup modes instead. A concrete choice is to consider the resolution model with line bundle vectors 
\begin{equation}
V_\alpha  = (0^2, \tfrac 12^6)(0^8)~, 
\end{equation} 
for all $\alpha = 1,\ldots, 16$. Since, again, the same line bundle vector is chosen on all exceptional divisors, the multiplicity operator reduces to~\eqref{eq:MultOpAllEqual}. The spectrum can be determined as before and is given in Table~\ref{tb:ResT4Z2StEmbSpecSpinorial}.

\begin{table} 
\[ \renewcommand{\arraystretch}{1.6}
\begin{array}{|c|c|c|c|c|} 
\hline 
\text{weight} & \mathsf{H}_V^2  & \mathsf{N} &  SU(2)_1\times E_6 \times E_8' & SU(2)_1 \times SU(2)_2\times SU(6) \times SO(16)' 
\\ \hline\hline   
\pm(1^2,0,0^5)(0^8) & 0 & -2_U & SU(2)_1~\text{gauge}  & SU(2)_1~\text{gauge}
\\  \hline
(0^2, \underline{1,-1,0^4})(0,^8) & 0 & -2_U & E_6~\text{gauge} & SU(6)~\text{gauge}
\\ 
(\underline{1,-1},0^6)(0,^8) & &  &  & SU(2)_2~\text{gauge}
\\ 
(\underline{ \tfrac 12,  -\tfrac 12}, \underline{-\tfrac 12^3, \tfrac 12^{3}})(0^8)  &  &  &  & (1,2,20)(1)~\text{gauge}
\\ \hline
\pm(\underline{\pm1,0},\underline{1,0^5})(0^8) & \tfrac 14 & 2_U & (2,27)(1)  & (2,2,6)(1) 
\\  
\pm (\tfrac 12^2, \underline{\tfrac 12^4, -\tfrac 12^{2}})(0^8)  &  &  &  & (2,1,15)(1)
\\  
\pm (-\tfrac 12^2,  \underline{\tfrac 12^4, -\tfrac 12^{2}})(0^8)  &  &  &  & 
\\ \hline  
\pm(0^2,\underline{1^2,0^4})(0^8) & 1 & 14 = 16_T -2_U & (1,27)(1) & (1,1,15)(1) 
\\  
\pm ( \underline{\tfrac 12, -\tfrac 12},  \underline{\tfrac 12^5, -\tfrac 12})(0^8) &  &  &  & (1,2,6)(1)
\\  \hline  
\pm ( \tfrac 12^2,  \tfrac 12^6)(0^8)  & \tfrac 94 & 34=32_T+2_U & (2,1)(1)  & (2,1,1)(1) 
\\ 
\pm ( -\tfrac 12^2,  \tfrac 12^6)(0^8)  &  & &   &
\\ \hline\hline  
\pm (0^8)(\underline{\pm1^2,0^6}) & 0 & -2_U & E_8~\text{gauge} & SO(16)~\text{gauge} 
\\
\pm (0^8) ( \underline{-\tfrac 12^e, \tfrac 12^{8-e}}) &  &  &  & (1,128)~\text{gauge}  
\\  \hline 
\end{array}
\]
\caption{
\label{tb:ResT4Z2StEmbSpecSpinorial}
The multiplicities of the $E_8\times E_8$ roots are indicated for the resolution model generated by identical spinorial blowup modes at all sixteen exceptional divisors. The states with a positive or a negative multiplicity form hyper or vector multiplets. The subscripts $U$ and $T$ indicate how these numbers can be used to interpret the corresponding states as untwisted or twisted, respectively. The final column gives the spectrum branched by the Wilson line. }
\end{table}

\subsubsection*{Matching with the Orbifold Spectrum}

The above resolution model can also be understood as the blowup of the orbifold standard embedding model. In this case blowup modes all have shifted momenta
\begin{equation}
P_{\text{sh},\alpha} = V_\alpha  = V + P~, 
\quad 
V= (\tfrac 12^2, 0^6)(0^8)
\quad\text{and}\quad 
P= (-\tfrac 12^2,\tfrac 12^6)(0^8)
\end{equation}
at all sixteen fixed points. They live on the (shifted) spinorial lattice of $SO(16)$ and part of the sixteen half--hyper multiplets $(1,56)(1)$. Switching on these blowup modes lead to the symmetry breaking: 
\begin{equation}
SU(2) \times E_7 \times E_8' \rightarrow SU(2) \times E_6 \times E_8' 
\end{equation}
In this process precisely the roots $\pm(0^2,\underline{1^2,0^4})(0^8)$ and 
$\pm(\underline{\tfrac 12,-\tfrac 12}, \underline{-\tfrac 12,\tfrac12^5})$ of the $(1,27)(1)$ are broken. This, again, corresponds to the computation $14 = 16_T - 2_U$, which can be understood by the super-Higgs effect where certain twisted states are ``eaten'' to form massive vector multiplets. These are states that arise from the 16 half-hyper multiplets $(1,56)(1)$. Under the symmetry breaking this branches to 
\begin{equation}
\tfrac 12(1,56)(1) \rightarrow (1,27)(1) + (1,1)(1)
\end{equation}
As before,  the states $(1,1)(1)$ are the blowup modes (BLW), which on the resolution are reinterpreted as 16 model dependent axions not contributing to the multiplicity operator.

The remaining 14 states  $(1,27)(1)$ after the Higgsing undergo a field redefinition when moving from the hyper multiplets on the orbifold to the states on the resolution: 
\begin{equation}
(1,27)(1)_\text{RES} = \text{BLW}^{-1}\cdot (1,27)(1)_\text{ORB} 
\end{equation}
Indeed, the corresponding weights can be matched exactly via:
\begin{equation}\renewcommand{\arraystretch}{1.6}
\begin{array}{ccccc}  
 \pm(0^2,\underline{1^2,0^4})(0^8)
 &  =  & 
 \mp \big[ -(0^2,\tfrac 12^6)(0^8) & + &  (0^2, \underline{-\tfrac 12^2,\tfrac 12^4}) \big]   
 \\
 \pm (\underline{\tfrac 12,-\tfrac 12},\underline{-\tfrac 12,\tfrac 12^{5}})(0^8) 
 &  =  & 
 \mp \big[ -(0^2,\tfrac 12^6)(0^8) & + & (\underline{-\tfrac 12,\tfrac 12}, \underline{1,0^5}) \big]   
\end{array}
\end{equation}
A similar field redefinition relate the doublet states on the orbifold to those on the resolutions: 
\begin{equation}
(2,1)(1)_\text{RES} = \text{BLW}^{-1}\cdot (2,1)(1)_\text{ORB}~,
\end{equation}
or in terms of the corresponding weights: 
\begin{equation}
\pm (\mp\tfrac 12^2,\tfrac 12^6)(0^8)   = \mp\big[ -(0^2,\tfrac 12^6)(0^8) + (\pm\tfrac 12^2, 0, 0^5) \big]
\end{equation}
This analysis shows that the spectrum on this orbifold resolution with spinorial blowup modes is the same as for the previous choice of bundle vectors corresponding to vectorial blowup modes. This can be seen explicitly by comparing columns three and four of Table~\ref{tb:ResT4Z2StEmbSpec} and Table~\ref{tb:ResT4Z2StEmbSpecSpinorial}. The effect of the Wilson line is very different however.

\subsection{Wilson line projected spinoral blowup model -- With Torsion} 

Since the spinorial blowup modes were precisely considered to avoid the issue that the blowup modes are projected out by the Wilson line on the additional torus if torsion is switched on, the case with torsion is discussed below. (The spinorial blowup model without torsion suffers from similar issues as the vectorial blowup model with torsion and is ignored in the following.)

As stressed on resolution geometries one has to make an educated guess how torsion between the Wilson line and the orbifold twist is implemented based on intuition derived from the orbifold description. Concretely, the spinorial blowup model is further compactified on a circle $S^1$ with a discrete Wilson line: 
\begin{equation}  
W = (0^7, 1)(1,0^7)~. 
\end{equation}
The Wilson line projection condition in the presence of torsion on this resolution is assumed to be implemented as follows: 
\begin{equation} 
W\cdot P \equiv 0~.
\end{equation}
The motivation for this from the orbifold description is, that for the twisted states, which feel the presence of the torsion phase, the relation between the $P_\text{sh}$ and $P$ involves $V_\alpha$ for which $W\cdot V_\alpha = \tfrac 12$ in this case. Hence, the effect of the torsion for the projection condition~\eqref{eq:PhaseProjection}  is compensated by the fact that the blowup mode itself is spinorial. Consequently, the gauge group on the blowup is: 
\begin{equation}
SU(2)_1\times SU(2)_2 \times SU(6) \times SO(16)'
\end{equation}
and the resulting five dimensional spectrum reads: 
\begin{equation}
2\, (2,2,6)(1) + 14\,(1,1,15)(1)~. 
\end{equation}

This is compatible with the orbifold spectrum with the Wilson line and the torsion phase.  Indeed, the spinorial blowup models induce the symmetry breaking: 
\begin{equation}
SO(12) \rightarrow SU(6)~. 
\end{equation}
The broken generators are $\pm(0^2,\underline{1^2,0^4})(0^8)$ correspond to two massive vector multiplets in the  $ (1,1,15)(1)$ representation. The charged orbifold spectrum branches as follows: 
\begin{equation}
\begin{array}{rcl}
(2,2,12)(1) &\rightarrow& 2\, (2,2,6)(1)~, 
\\[2ex]
\tfrac 12 (1,1,32)(1) &\rightarrow& (1,1,1)(1) + (1,1,15)(1)~. 
\end{array}
\end{equation}
The sixteen singlets $(1,1,1)(1)$ are the sixteen blowup modes and appear on the blowup as axions. Two of the $(1,1,15)(1)$ pair up with the broken generators to form the two massive vector multiplets. This leaves fourteen $(1,1,15)(1)$ in the massless charged spectrum.

\begin{table} 
\[
\begin{array}{|l||c|c|}
\hline 
\text{Torsion Phase } (\epsilon) & \text{Without } (\epsilon = 0) & \text{With } (\epsilon = 1)  
\\ \hline\hline
\text{Orbifold} && \\ 
\text{Gauge Group} & 
SU(2)_1 \times SU(2)_2 \times SO(12) \times SO(16)' & SU(2)_1 \times SU(2)_2 \times SO(12) \times SO(16)'
\\ 
\text{Spectrum} & 
(2,2,12)(1) + 16\, \tfrac 12(1,2,12)(1) & (2,2,12)(1) + 16\, \tfrac 12(1,1,32)(1) 
\\ & 
+ 32\, (2,1,1)(1) + 4 (1,1,1)(1) & + 4 (1,1,1)(1)
\\ \hline\hline 
\text{Blowup} && \\ 
\text{Modes } P_{\text{sh},\alpha} =V_\alpha & 
(\tfrac 12,-\tfrac 12,1,0^5)(0^8)  & (0^2, \tfrac 12^6)(0^8)
\\ 
\text{Gauge Group}  &  
SU(2) \times SO(10) \times SO(16)' & SU(2)_1 \times SU(2)_2 \times SU(6) \times SO(16)'
\\
\text{Spectrum} & 
2\, (2,10)(1) + 36\, (2,1)(1) & 2\,(2,2,6)(1) + 14\,(1,1,15)(1)
\\ 
& + 14\, (1,10)(1) + 14\, (1,1)(1)  & 
\\\hline 
\end{array}
\]
\caption{ \label{tb:SVDOrbRes} 
This table summarises how a spinor--vector duality is visible in orbifold and resolution models. Since the resolutions depend on the choice of blowup modes, their gauge groups and therefore their spectra make this duality less apparent. 
}
\end{table}

\subsection{Spinor--Vector Duality on Resolutions}

In this final subsection some possible lessons for the realisation of spinor--vector dualities on orbifold resolutions and smooth geometries in general are discussed based on the results of the previous subsections. 

Like in free fermionic models, also on orbifolds one expect spinor--vector dualities to be present and easily identifiable. Both descriptions have an underlying worldsheet structure and can be encoded in (one--loop) string partition functions in which additional torsion phases may be present. The dictionary between free fermionic and orbifold models developed in~\cite{Athanasopoulos:2016aws} may be used to relate these description in all fine print. 

Moving from the orbifold point to smooth resolutions blowup modes have to be selected. These are twisted states which develop VEVs inducing the blowup of the orbifold singularities. Since the Wilson lines on (additional) cycles lead to projections of the twisted spectrum with or without torsion, the selected blowup modes may not be in the spectrum anymore, which leads to various complications. A prime one being that the choice of the torsion phase affects which blowup modes are available.

In the particular cases considered here of orbifold and resolution models of $T^4/\Intr_2 \times S^1$ with a Wilson line, the  spinor--vector duality mapping is summarised in Table~\ref{tb:SVDOrbRes}. On the orbifold the spinor--vector duality can clearly be seen: the model without torsion contains sixteen additional $SO(12)$ vector which are also $SU(2)$ doublets but no $SO(12)$ spinors, while the model with torsion has sixteen $SO(12)$ spinors but the $SO(12)$ vectors are absent. On the resulting resolutions using blowup modes indicated in the table the picture is far less transparant because the gauge groups in both cases are different. But the important characteristics of the spinor--vector duality can still be identified: In the resolution model without torsion in total eighteen vectors of $SO(10)$ are present while in the model with torsion there are eight vectors, $(6)$--plets, and fourteen anti--symmetric tensors, the $(15)$--plets, of $SU(6)$ present. These $(15)$--plets can only arise from the branching of the spinorial representation of $SO(12)$. Hence, the spectra on the resolutions still exhibit properties associated to the spinor--vector duality, albeit in some disguise.

\section{Conclusion}
\label{sc:Conclusion}

\subsection*{Summary}

String theory provides a self--consistent framework for the synthesis of all the matter and interactions seen in observational data.  The $\Intr_2\times \Intr_2$ orbifold, in its fermionic incarnation as well as the bosonic, gave rise to a multitude of phenomenological three generation models with different unbroken $SO(10)$ subgroups(see {\it e.g.}\ \cite{Antoniadis1989,Faraggi:1989ka,Antoniadis1990,Faraggi:1991jr,Faraggi1992a,Blaszczyk:2009in,Faraggi:2019drl}  and references therein). Spinor--vector duality plays a role in some of these constructions as well. Of particular note is the $Z^\prime$ model of ref.~\cite{Faraggi:2014ica,Faraggi:2018pit} in which self--duality under the spinor--vector duality is instrumental to obtaining a three generation model with an extra $U(1)$ gauge symmetry, which is family universal and with the standard $E_6$ embedding of the $Z^\prime$ charges. It was argued in ref.~\cite{Faraggi:2018pit} that existence of light sterile neutrinos mandates the existence of such an extra symmetry under which the sterile neutrinos are chiral and which remains unbroken down to low energy scales.

The main aim of this paper was to study the spinor--vector duality on smooth geometries. Inspired by the models presented in~\cite{Faraggi:2011aw} the orbifold $T^4/\Intr_2$ with an additional circle with a Wilson line, is considered. This Wilson line distinguishes between integral and half--integral weights in the string spectrum. As to be expected from that paper depending on a generalized torsion phase between the orbifold twist and the Wilson line, the resulting five dimensional models indeed exhibit a spinor--vector duality. 

Using standard resolution techniques the blowup of the orbifold $T^4/\Intr_2$ was constructed. Since, this orbifold by itself leads to a six dimensional model, the full massless spectrum on the resolution can be determined with the help of the multiplicity operator and was shown to match completely with the orbifold spectrum upon taking field redefinitions involving the blowup modes into account. After that the effect of the Wilson line on the additional circle was considered in the resolution setting. In the resolution model where the blowup modes are all vectorial, this resulted in a projection of the massless spectrum consistent with the expectations from the orbifold theory. 

On smooth geometries the interpretation and implementation of the torsion phase between the orbifold twist and the Wilson line is obscured as there is no notion of the former. Since the smooth geometry in the present paper was obtained as an orbifold resolution, the effect of the generalised torsion phase on the blowup could be conjectured to act as expected from the orbifold theory. But proceeding in this way led to an inconsistent spectrum. The reason for this could be traced to the fact that, because of the torsion phase the vectorial twisted states were projected out, but precisely those were used to generate the blowup. To overcome this problem, a second resolution model was considered, where spinorial twisted states were used as blowup modes instead. The effect of the Wilson line with the torsion phase switched on is to keep them in the orbifold theory and the resolution spectrum made sense again. However, because the spinorial blowup modes led to a further symmetry breaking, the gauge group of interest was no longer $SO(10)$ but rather $SU(5)$. Table~\ref{tb:SVDOrbRes} collects the uncovered details of the spinor--vector duality on the orbifold and its resolution. 

To summarise, an example of the spinor--vector duality could be realised on a smooth resolution, but the picture of the duality is more subtle as the gauge groups of the dual models are not the same. The underlying reason was that the available blowup modes with the torsion phase switched on or off are complementary, so that different blowup modes are needed to be selected depending on the torsion choice. We expect this feature to be generic as long as the generalised torsion involves the orbifold twist, since the projection of the twisted states (the candidate blowup modes) then depends on the choice of the torsion phase. Of course, there may be other ways, that a spinor--vector duality can be induced on smooth compactifications.

\subsection*{Outlook}

One complication encountered in this work was how to implement generalised torsion phases on smooth geometries. In particular, in effective supergravity compactifications it is not clear how the generalised GSO phases of string theory should be taken into account. It has been argued that certain forms of discrete torsion can be understood as a group action on the $B$--field~\cite{Sharpe:1999pv,Sharpe:1999xw,Sharpe:2000ki,Sharpe:2000tw}. This description might help to develop a deeper understanding of generalised GSO projections on smooth geometries. 

In this paper we considered the resolution of the simple orbifold $T^4/\Intr_2$ as the starting point of our analysis of the spinor--vector duality on smooth geometries. In a future publication we plan to investigate more complicated orbifold resolutions, like, in particular, the resolutions of $T^6/\Intr_2\times \Intr_2$.


\bibliographystyle{paper}
{\small
\bibliography{paper}
}
\end{document}